\documentclass[debug,overfull]{epl}
\newcommand{\dt}{\mbox{$d_t$}}
\newcommand{\de}{\mbox{$d_e$}}
\newcommand{\Nt}{\mbox{$N_t$}}
\newcommand{\Ne}{\mbox{$N_e$}}
\newcommand{\Nxi}{\mbox{$N_{\xi}$}}
\newcommand{\De}{\mbox{$D_e$}}
\newcommand{\nud}{\mbox{$\nu_0$}}
\newcommand{\md}{\mbox{$m_0$}}
\newcommand{\bd}{\mbox{$b_0$}}
\newcommand{\bs}{\mbox{$b$}}

\title{On two intrinsic length scales in polymer physics:\\
topological constraints {\em vs.} entanglement length}

\author{
M. M\"uller\inst{1} \and 
J.P. Wittmer\inst{2}\thanks{E-mail: \email{jwittmer@dpm.univ-lyon1.fr}}
\and
J.-L. Barrat\inst{2}}
\shortauthor{M. M\"uller \etal}
\shorttitle{Topological constraints vs. entanglement length}

\institute{
\inst{1}Institut f\"ur Physik, Johannes Gutenberg-Universit\"at,
Staudinger Weg 7, D-55099 Mainz, Germany.\\
\inst{2}D\'epartement de Physique des Mat\'eriaux,
Universit\'e Claude Bernard and CNRS, 69622 Villeurbanne Cedex, France.
}

\rec{}{}

\begin{document}

\maketitle

\begin{abstract}
The interplay of topological constraints, excluded volume interactions, 
persistence length and dynamical entanglement length in solutions and 
melts of linear chains and ring polymers is investigated by means 
of kinetic Monte Carlo simulations of a 
three dimensional lattice model.
In unknotted and unconcatenated rings,
topological constraints manifest themselves in the static properties above
a typical length scale  $\dt \sim 1/\sqrt{l\phi}$ 
($\phi$ being the volume fraction, $l$ the mean bond length). 

Although one might expect that the same topological length will play a
role in the dynamics of entangled polymers, we show that this is not the case.
Instead,  a different intrinsic length \de, 
which scales like excluded volume blob size $\xi$, 
governs the scaling of the dynamical properties of both linear chains and rings. 
In contrast to \dt, \de\ has a strong dependence on the chain stiffness.
The latter property enables us to study the full 
crossover scaling in dynamical properties, 
up to strongly entangled polymers.
In agreement with experiment the scaling functions of both architectures
are found to be very similar.
\end{abstract}

{\em Introduction.}
Topological constraints (TC) are commonly invoked in polymer theory
to explain the dynamical behaviour of large molecular weight chains. 
In the reptation theory \cite{degennesbook,edwardsbook,lrp,KG95},
the long range TC are assumed to be responsible on a local scale for the 
conjectured anisotropic motion of chain segments,
which are not allowed to cross neighboring chains. 
Obviously, in systems of linear chains this non-crossing constraint does
not provide TC in a strict mathematical sense \cite{edwardsbook}. 
Strict topological constraints, however, do exist in systems 
of unconcatenated and unknotted ring polymers, where they strongly
influence the static properties \cite{CD86,ORD94,MWC96,MWC00}.
In this case,  the TC are created by the non-crossing requirement 
and, as has been shown recently\cite{MWC00}, 
a mass independent length scale \dt. It is, hence, natural to ask
whether  \dt\ provides the conjectured length scale,
i.e. the "tube diameter" $d_e$ \cite{edwardsbook}, 
for the dynamics of linear chains and rings. 
A link between the two concepts was indeed suggested by the fathers of the 
reptation concept, de~Gennes\cite{degennesbook} and Edwards\cite{edwardsbook}.
This idea prompted the present computational study, which extends our previous
investigation on semiflexible rings on linear polymers.
Concentrating on the scaling analysis of global quantities
such as the diffusion constant and the radius of gyration, 
we demonstrate that the static and dynamic
problems involve different length scales, with different 
dependence on chain stiffness and on density.
The conjecture $\de \approx \dt$ is therefore incorrect,
and the simpler proposal $\de \approx \xi$ \cite{edwardsbook,KG95}, 
where $\xi$ is the correlation blob size, accounts much better for 
the scaling behaviour.

\begin{figure}
\twofigures[scale=0.4]{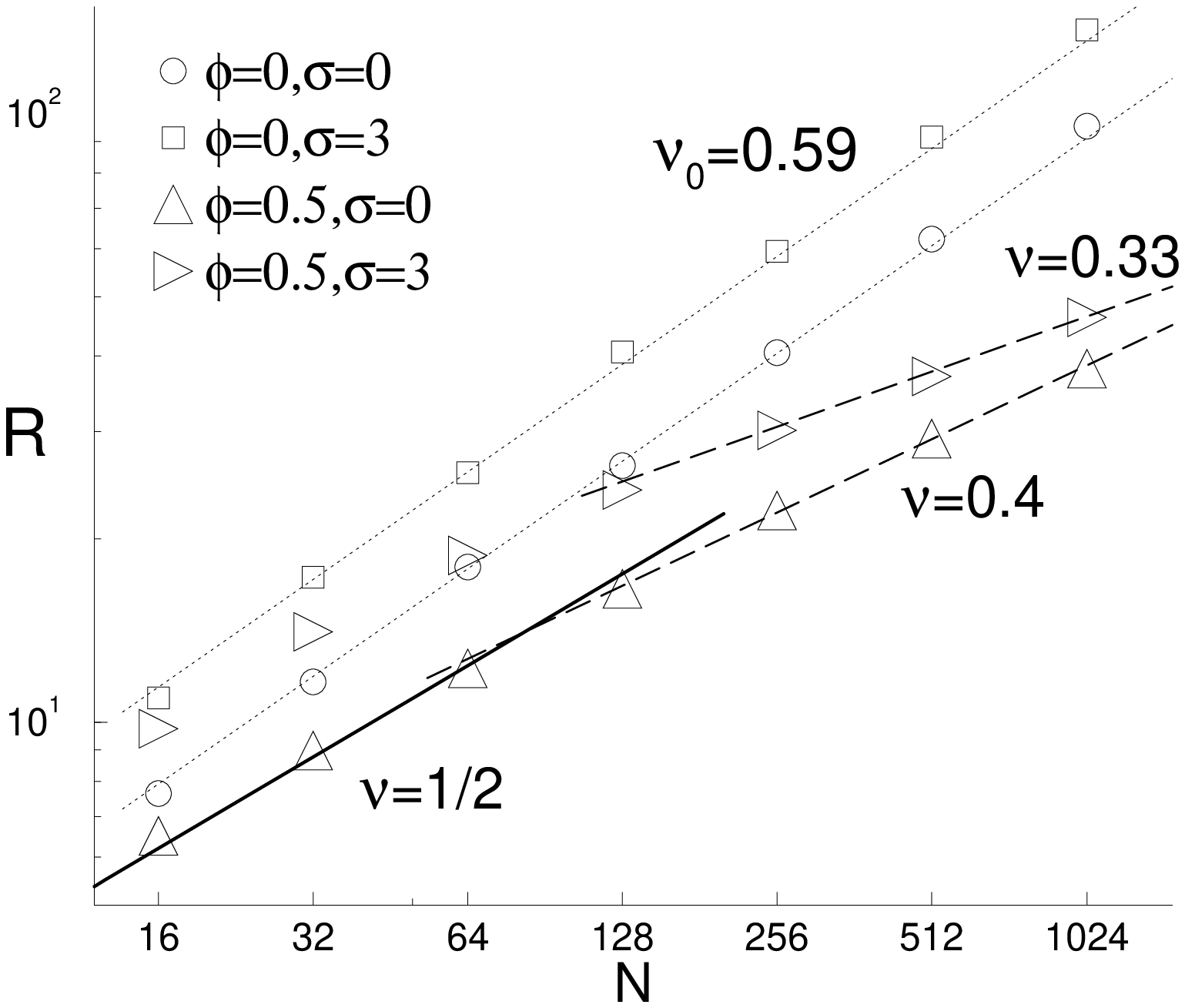}{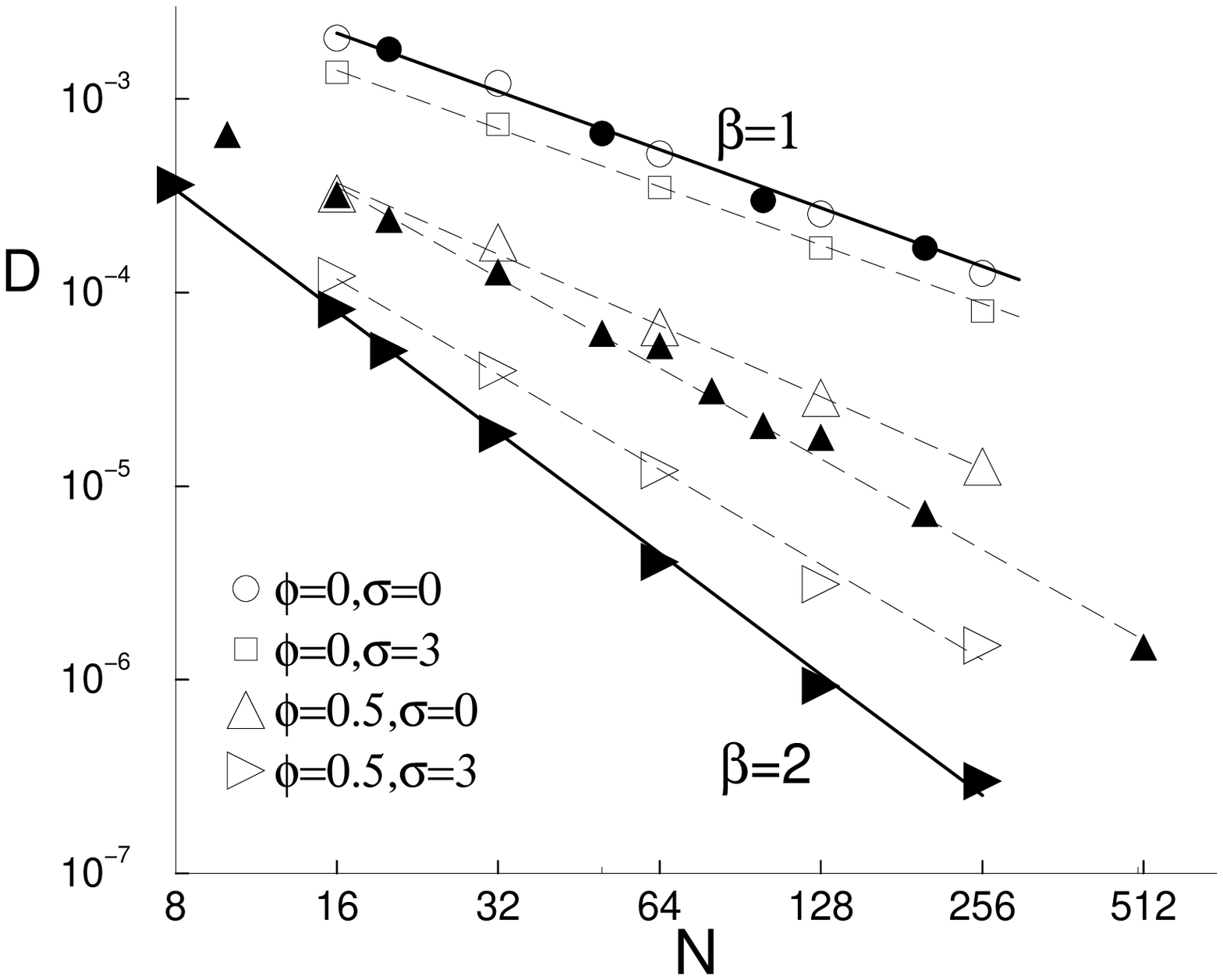}
\caption{
Ring size $R=\langle \vec{R}_0-\vec{R}_{0+N/2}\rangle^{1/2}$ versus 
chain length $N$ for dilute ($\phi\rightarrow 0$) and molten rings. 
The dilute rings are characterized by the same Flory exponent 
$\nu=\nud\approx 0.59$ as their linear counterparts (not included).
In the dense limit short rings form Gaussian chains of blobs 
($\nu=1/2$: solid line). 
With chain overlap, i.e. with increasing molecular mass, volume fraction and 
(most importantly) stiffness, the rings become more and more compact.
\label{figRN}}
\caption{Diffusion coefficient $D(N,\phi,\sigma)$ versus chain length $N$. 
We consider flexible ($\sigma=0$) and semiflexible ($\sigma=3$)
linear chains (full symbols) and rings (empty symbols) in the dilute 
and the melt limit. 
In the dilute limit both architectures have the same Rouse-like dynamics
(upper bold line).
Contrasting to this the dynamics at high overlap depends strongly on both 
architecture and stiffness. 
Effective power law exponents $D\propto N^{-\beta}$ are indicated in the figure
(dashed lines); $\beta$ increases with $N$,$\phi$ and $\sigma$.
Rings are always faster than corresponding linear chains and have smaller
apparent exponents. The $\beta \approx 2$-slope --- consistent with the 
reptation prediction\protect\cite{degennesbook,edwardsbook} --- 
is observed over more than one order of magnitude for semiflexible ($\sigma=3$)
linear chains (lower bold line).
\label{figDN}}
\end{figure}

{\em Computational method and raw data.}
As in our previous studies\cite{MWC00,WPB92}, we use Monte-Carlo simulations
of an extensively studied lattice model, the bond fluctuation model\cite{bfm}. 
This model, in which the dynamics proceeds by local jump attempts,
is thought to 
describe well the polymer dynamics on Langevin equation level\cite{paul}.
In order to tune the persistence length, we impose a simple intramolecular
potential which favors straight bond angles $E(\theta)=\sigma \cos(\theta)$
where $\theta$ denotes the complementary angle between two successive bonds.
The stiffness $\sigma$ (in the range $\sigma=0,1,2$ and $3$) is an energy scale,
expressed here in units of the thermal energy  $k_BT$
\cite{MWC00,WPB92}.  

In fig.~\ref{figRN} and fig.~\ref{figDN} we present two sets of raw data from 
our simulations  --- the ring size $R$ and the self-diffusion constant $D$ for
linear chains and rings as functions of their molecular mass $N$. 
We have sampled a broad range of volume fractions (of occupied lattice sites) 
$\phi$ from the dilute up to the melt\cite{paul} limit ($\phi\approx 1/2$) 
with chain masses up to $N=1024$ 
%%%(as indicated in the figures)
in order to vary the chain overlap sufficiently to test the proposed scaling
scenarios. 

In the bond fluctuation model the bond length $l$ is not, strictly speaking, 
a constant, since a bond can be represented by lattice vectors
of different lengths. The average bond length $l$, however, depends very
weakly on stiffness and density, and will be considered as constant.
The data for dilute chains can be used to define
the effective or statistical segment length $\bd(\sigma)$, 
obtained from fitting the radius of gyration of 
asymptotically long dilute linear chains with $R_{gyr}=\bd(\sigma) N^{\nud}$ 
where $\nud\approx 3/5$ is the swollen chain exponent. $\bd(\sigma)$ increases 
with stiffness up to 50\%.
Similarly, linear chains above the overlap concentration are Gaussian,
and obey the scaling $R_{gyr}=\bs(\phi,\sigma)N^{1/2}$.  
Local dynamics is characterized by a monomer 
mobility $m(\phi,\sigma)$, obtained from the mean-square displacements (MSD)
of center monomers at short times \cite{WPB92}. 
Again, this quantity turns out to be 
relatively insensitive to stiffness.
In the dilute limit
it may be equally be obtained from the center-of-mass motion. 
As shown in fig.~\ref{figDN} we find $D=\md(\sigma)/N$
for both architectures, with a mobility \md\ roughly proportional 
to the acceptance rate. 
At higher densities we find $m(\phi,\sigma)\approx\md(\sigma)\tilde{m}(\phi)$
with $\tilde{m}(0)=1$, i.e. density and rigidity effects decouple.

While static and dynamic properties of both architectures resemble each 
other in the dilute limit and at moderate overlap,
where both linear chains and rings form Gaussian chains of blobs -- 
bold line in fig.~\ref{figRN} -- 
marked differences appear at high chain overlap,
as seen in figures \ref{figRN} and \ref{figDN}. 
The local slopes in  these figures, 
$\nu = d\log R/d\log N $ and $\beta= -d\log D/d\log N$ 
depend continuously on $N$, indicating that our data 
extensively cover the crossover regions.
With increased chain overlap the rings become more compact 
($1/\nu\rightarrow 3$) and their dynamics more entangled-like,
i.e. the diffusion exponent $\beta$ approaches 2 \cite{com:entangled}. 
The stiffness dependence is very strong, especially for the diffusion constant.
That the reptation prediction $\beta=2$ for linear chains
\cite{degennesbook,edwardsbook} is indeed seen so clearly,
is one of the important results of this work; 
this is in agreement with a very recent MD simulation \cite{Faller}.
Note that rings are always {\em faster} than their linear counterparts.
%in contradiction to a somewhat naive reading of the reptation picture.

Obviously, the slopes are only effective exponents which characterize the crossover 
between different regimes and do not capture asymptotic behaviour 
(e.g., the two slopes indicated in fig.~\ref{figRN}
would then intercept which is unphysical).
The task is then to understand the effects on both architectures of increased 
chain overlap --- specifically, the strong stiffness dependence ---
and to elucidate the underlying crossover scaling.

{\em Two intrinsic length scales and scaling attempts.}
There are two\cite{com:Fedderlength} natural intrinsic length scales --- 
both chain length independent ---
which have been used to characterize the interactions in strongly overlapping 
polymer solutions and melts\cite{degennesbook}:
the 'diameter' $\rho(\phi)=1/\sqrt{l\phi}$ 
and the `blob' size $\xi(\phi,\bd) \approx \bd \Nxi^{\nud}$
where $\Nxi \approx (\bd^3\phi)^{-1/(3\nud-1)}$ is the number of monomers
contained in the blob. 
Note that in contrast to $\rho$ the blob size $\xi\propto \bd^{-1.3}$ is
strongly stiffness dependent.
The prefactors of $\xi$ and $\Nxi$ can be estimated independently
from the static structure factor or, equivalently, from 
$\bs(\phi,\sigma)=\xi/\Nxi^{1/2}$\cite{MWC00,WPB92}. 

\begin{figure}
\onefigure[scale=0.60]{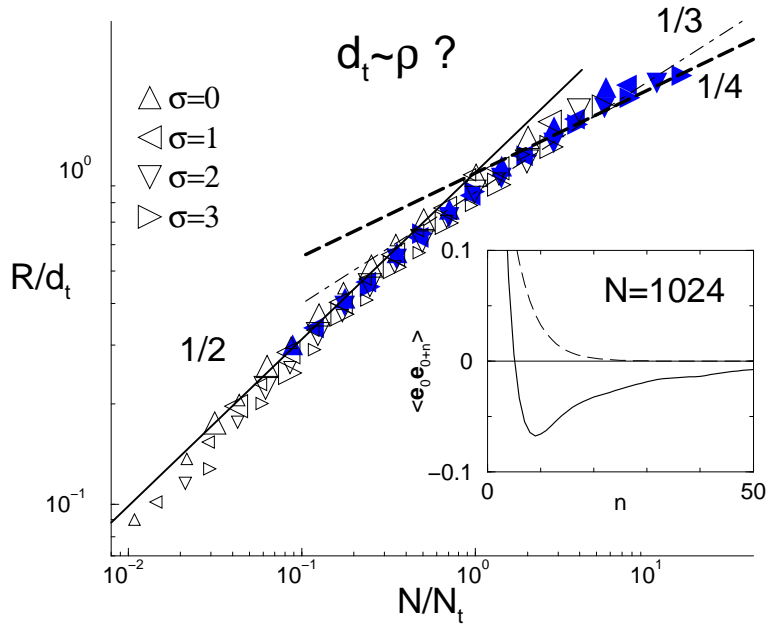}
\caption{Data collapse of the reduced ring size $R/\dt$ versus $N/\Nt$
for rings at three representative densities 
($\phi=1/2$: full symbols; $\phi=1/8$: large, thin symbols; $\phi=1/32$: small symbols)
at different $\sigma$ as indicated in the figure.
The successful scaling attempt supposes $\dt \approx 25 \rho$ and 
self-consistently (full line on the left) 
Gaussian chains of blobs at small distances, 
i.e. $\Nt\approx 405 (\rho/\bs(\phi,\sigma))^2$.
The slope $\nu=1/4$ for the predicted
%\protect\cite{ORD94} 
LA asymptotic 
behaviour for larger chains is indicated on the right (dashed) along
with the apparent $\nu=1/3$-slope (dashed-dotted) from fig.~\protect\ref{figRN} 
which fits a broader range of data.
Inset: The bond-bond correlation function 
$\langle \vec{e}_0 \cdot \vec{e}_{0+n} \rangle$
shows striking anti-correlations for rings (bold line) at large overlap
($N=1024$,$\phi=0.5$,$\sigma=3$) in contrast to linear chains (dashed).
\label{figRscal}}
\end{figure}

\begin{figure}
\onefigure[scale=0.60]{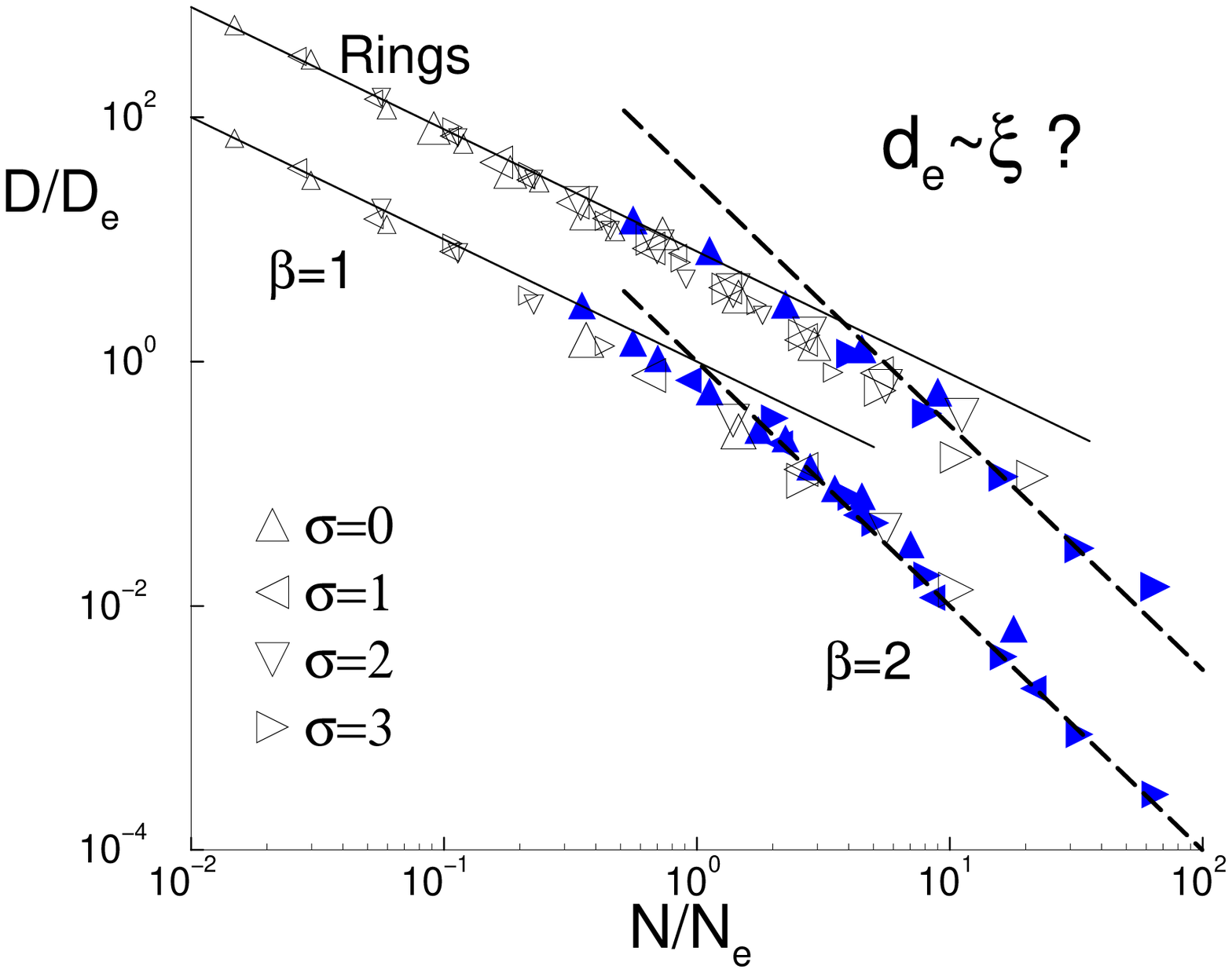}
\caption{Scaling attempt for diffusion constants of linear chains
and rings (shifted upwards for clarity) supposing $\de \sim \xi$.
$D/\De$ is plotted versus $N/\Ne$ where we use $\De = m(\phi,\sigma)/\Ne$
and $\Ne = 32.8 (\bd^3\phi)^{-1/(3\nud-1)}$.
Data for four rigidities (as indicated in the figure) and three densities
(as in fig.~\ref{figRscal}; same symbols) have been included. 
The data collapse was achieved without any free parameter!
The linear chain data follow the reptation 
prediction for nearly two orders of magnitude.
The prefactor of \Ne \ was defined such that both linear chain asymptotes 
(bold and dashed lines) intercept at $D/\De=N/\Ne=1$. 
\label{figDscal}}
\end{figure}

We demonstrate now that $\rho$ determines \dt\ (Fig.~\ref{figRscal}) 
and that the polymer dynamics is indeed characterized 
by {\em one} length scale $\de\sim\xi$ (Fig.~\ref{figDscal}).
Note that the figures include data from very different densities and rigidities 
and that no shift or fit parameter has been used.  
It is also important to remark that data points  
may be found in the different parts of the scaling functions, 
irrespective of their density or stiffness. 
This shows that the true intrinsic parameter is neither $\phi$ nor $\sigma$,
but the proposed scaling variable.

First, we have verified that the traditional scaling $R/\xi$ versus 
$N/\Nxi$\cite{degennesbook} works for semiflexible linear chains (not shown). 
Surprisingly, it does not work for rings \cite{MWC00}.  
This is an indication of the influence of TC on static properties.  
In fig.~\ref{figRscal}, we show that a proper scaling of the 
radius of gyration for rings is obtained by using the scaling variables
$R/d_t$ and $N/N_t$, where  $\dt \approx 25/\sqrt{l\phi}$. The number $N_t$ 
of monomers 
between topological obstacles \Nt\ used in fig.~\ref{figRscal}
is obtained from $\dt \approx b(\phi,\sigma) \Nt^{1/2}$ where we assume 
Gaussian 
statistics at distances smaller \dt. 
This is motivated by fig.~\ref{figRN} 
% (bold line) 
and is consistent with the scaling (left side of fig.~\ref{figRscal}).
From this successful scaling collapse we may conclude that the length 
generated by the TC is $\dt \approx 25/\sqrt{l\phi}$.
This yields $\dt\approx 22$ for $\phi=0.5$ and 
$\Nt(\sigma=0) \approx 173$ and $\Nt(\sigma=3) \approx 63$.
This length  is both mass and stiffness independent. 
The prefactor is somewhat arbitrary and depends on the asymptotic slope on the
right which is numerically unclear (see below). 

We turn now to the scaling of the diffusion data, and demonstrate,
 that polymer dynamics is indeed characterized 
by {\em one} length scale \de \ which scales like the blob size $\xi$. 
In fig.~\ref{figDscal}, the mass has been rescaled by 
a number  \Ne\  assumed to be proportional to the number of monomers per blob \Nxi.
The scale for the vertical axis is 
given by  $\De = m(\phi,\sigma)/\Ne$, 
i.e. we assume Rouse-like behaviour for weakly overlapping chains.
For small $N$ all data collapse irrespective of architecture and stiffness
on the asymptotic slope with $\beta=1$. 
This shows that the definition of \De\ is self-consistent and also 
that the mobility correction $m(\phi,\sigma)$ was chosen appropriately.
For $N\gg \Ne$ we find that the linear chain data follow the reptation 
prediction $D/\De = (\Ne/N)^2$ for nearly two orders in magnitude!
From the known blob size and the intercept of both asymptotes one estimates 
$\Ne/\Nxi \approx 15$
% and $\de/\xi \approx 4$ 
for the linear chains.
%which is comparable, but somewhat smaller than other recent estimates \cite{lrp}.
%
Note that our scaling is consistent with the traditional density
crossover scaling assumption \cite{edwardsbook} and recent experiments 
on linear chains\cite{Graessley,com:Fedderlength}.  
Also, the present scaling, in contrast to that in Ref.~\cite{Faller},  
does not involve any arbitrary shift parameters.

The universal functions for the dynamical properties
of both architectures are 
%very 
similar.
In agreement with experiment\cite{lrp,Lutz} ring dynamics becomes
entangled-like at slightly higher chain length. 
This is no surprise and may be attributed to their smaller size
\cite{lrp}.
%\cite{lrp,inpreparation}.
However, the asymptotic behaviour is less obvious.
The best fit for the available data $\beta\approx 1.7$, 
may be regarded as lower bound. 
Motivated by recent experiments\cite{Lutz} which show the same asymptotic behaviour 
for both architectures we have also fitted the ring data with $\beta=2$
yielding a reasonable agreement  (upper dashed line in fig.~\ref{figDscal}).
From the intercept of the unentangled and entangled asymptote one
concludes that the ring crossover occurs roughly at $3-4$ times larger
chain length than in their linear counterparts --- 
again in agreement with experiment\cite{Lutz}.
This finally explains why the $\beta$ exponents of rings at given 
$\phi$ and $\sigma$ are smaller.

{\em Interpretation of scalings results and new queries.}
Both presented scaling plots are consistent with two eminent 
mean-field descriptions in polymer physics 
--- the {\em lattice animal} (LA)
proposal for dense rings\cite{CD86,ORD94,MWC00} and the reptation concept
for polymer dynamics. They provide additional information in that
they identify the phenomenological length scales implicit in both theories.

Our compact rings at high chain overlap behave like LA in a self-consistent
network imposed by the TC of neighboring rings, i.e. they may be described on 
the same footing as isolated rings
in a network of fixed topological obstacles of density $1/\dt^3$ \cite{ORD94}.
If the ring size $R$ becomes larger than $d_t$ the rings are forced
to retrace their paths and the fractal dimension $1/\nu$ becomes that
of a strongly branched object. Within the single chain model it is
known that the rings are characterized by the exponent $\nu=1/4$ 
within an intermediate, but broad chain length window \cite{ORD94}. 
It is this suggestion we have followed in fig.~\ref{figRscal}
to fix the prefactor of $d_t$.
Note that the rings presented in fig.~\ref{figRscal} are strongly overlapping,
with $R$ much larger than the distance between chains 
(see \cite{MWC00}, fig.7), and that
we refer to `compactness' in the scaling sense, $\nu\leq 1/3$.

The effective topological interactions can be visualized directly with
the bond-bond vector correlation function 
% $\langle e_i \cdot e_{i+n} \rangle_i$
shown in the inset of fig.\ref{figRscal}. This function 
has a pronounced anti-correlation. The negative minimum
indicates that the ring has to 
fold back after $n\approx 10$ monomers to retrace its path. 
The position and the depth of the dip increase with $\phi$ and $\sigma$,
but are chain length independent  which
underlines the $N$-independence of \dt.
The $\dt\sim\rho$ relation reflects the topological origin of the interaction.
Indeed $\rho$ characterizes the volume per chain segment 
%$l \rho^2$ 
which is {\em invariant} with regard to chain conformation, chain length and stiffness. 
The advantage in using stiff chains is related to the  fact that $\dt$
does not depend on stiffness:
semiflexible rings `waste' fewer monomers on short length scales, 
may explore larger distances and become more compact. 
Tuning the stiffness  provides therefore  a much more efficient 
route to increase the chain overlap,
required for the mean-field picture to hold,
than a chain length variation \cite{MWC00}.
Confirming older\cite{WPB92} and more recent simulations\cite{MWC00,Faller}
we find that increasing the stiffness is an even more efficient trick for 
exploring the strongly entangled dynamics.
This is now rationalized by the scaling $\de\sim\xi$ in fig.~\ref{figDscal}
which implies that the strong stiffness dependence of the 
number of entanglements per chain $N/\Ne \propto \bd^{3.9}$ 
\cite{com:stiffness}.
We emphasize, however,  that we have {\em not} demonstrated in this study the 
reptation proposal.
The $\beta=2$ asymptote of fig.~\ref{figDscal} is not sufficient to
claim anisotropic motion of the chains along their contour\cite{lrp}
which requires, e.g., a more detailed analysis of the MSD.
This must take into account the non-asymptotic behaviour in the 
MSD generated by tube renewal\cite{Schaefer} and 
constraint release effects.
%%%\cite{inpreparation}.
The latter mechanism is likely to be the dominant relaxation mode 
for our entangled rings.
%%%\cite{ORD94,inpreparation}.
What has been demonstrated here in a computational 
study covering, for the first time, several orders of magnitude is
(i)~the existence of two distinct dynamical regimes, both
in linear chains (consistent with \cite{KG95,WPB92,paul,Faller})
and ring polymers,
(ii)~that the dynamical crossover is characterized by {\em one} length scale only,
(iii)~that this length scale is the size of the excluded volume blob,
i.e. a properly defined static quantity \cite{com:Rousemelt},
(iv)~that, surprisingly, the crossover scaling does not depend on the architecture.
However, the finding $\dt \gg \de$ for the strongly overlapping chains
in our simulations may explain naturally why the length scale \dt \ 
appears to play no role in the dynamics of rings of experimental\cite{Lutz} 
and computational relevance.
%%%\cite{inpreparation}.

{\em Summary.}  In this letter, 
we have elucidated,  by means of a numerical
experiment, the {\em scaling} of static and dynamic properties of strongly 
entangled linear chains and  rings 
with regard to density and  stiffness variation. Our results indicate that topological
constraints manifest themselves differently in the statics and
in the dynamics. The topological length which governs the crossover in 
the statics of rings is essentially independent of stiffness.  
This contrasts with the
strong stiffness dependence of the dynamical entanglement length, which governs
the crossover to reptation-like behaviour.  Although the high molecular weight
diffusion exponent agrees with the reptation prediction, 
we have not presented rigorous proofs for the anisotropic motion
of polymer chains. A very puzzling question in this context is the similarity
between the scaling functions for linear chain and ring dynamics
which is the focus of forthcoming work.
%This and technical issues not presented here will be considered in a 
%future publication \cite{inpreparation}.

\acknowledgments
We thank M.E.~Cates, K.~Binder, W.~Paul
and J.~Baschnagel for stimulating discussions. 
We are grateful to T.~Kreer and J.~Baschnagel for communication
of the diffusion constant $D(N=512,\phi=0.5,\sigma=0)$
prior to publication. Generous grants of CPU time on the CRAY T3E
computers at the HLR Stuttgart and the IDRIS Saclay as well as 
financial support by the ESF under SIMU grant are gratefully acknowledged.

%%%%%%%%%%%%%%%%%%%%%%%%%%%%%%%%%%%%%%%%%%%%%%%%%%%%%%%%%%%%%%%%%%%%%%%%%%%%%%%


\begin{thebibliography}{0}

\bibitem{degennesbook} 
\Name{De Gennes P.-G.}
\Book{Scaling Concepts in Polymer Physics}
\Publ{Cornell University, Ithaca, N.Y.}\Year{1979}.

\bibitem{edwardsbook}
\Name{Doi M. \and Edwards S. F.}
\Book{The Theory of Polymer Dynamics}
\Publ{Clarendon Press, Oxford}
\Year{1986}.

\bibitem{lrp}
\Name{Lodge T. P., Rotstein N. A. \and Prager S.}
%{\em Dynamics of entangled polymer liquids: Do linear chains reptate?}
\REVIEW{Advances in Chemical Physics}{79}{1990}{1}.

\bibitem{KG95}
\Name{Kremer K. \and Grest G.} 
\Book{Monte Carlo and Molecular Dynamics Simulations in Polymer Science}
\Publ{Oxford University Press}\Editor{K. Binder}\Year{1995};
\Name{P\"utz M. \etal}\REVIEW{Europhys. Lett.}{49}{2000}{735}.

\bibitem{CD86}
\Name{Cates M. E. \and Deutsch J. M.}
\REVIEW{J.~de Physique}{47}{1986}{2121}.

\bibitem{ORD94}
\Name{Obukhov S. P., Rubinstein M. \and Duke T.}
\REVIEW{Phys.Rev.Lett.}{73}{1994}{1263}.

\bibitem{MWC96}
\Name{M\"uller M, Wittmer J.P. \and Cates M. E.}
\REVIEW{Phys.Rev.E}{53}{1996}{5063}.

\bibitem{MWC00}
\Name{M\"uller M., Wittmer J. P. \and Cates M. E.} 
\REVIEW{Phys.Rev.E}{61}{2000}{4078}.


\bibitem{WPB92}
\Name{Wittmer J., Paul W. \and Binder K.}
\REVIEW{Macromolecules}{25}{1992}{7211}.

\bibitem{bfm}
\Name{Carmesin I. \and Kremer K.}
\REVIEW{Macromolecules}{21}{1988}{2819}.

\bibitem{paul}
\Name{Paul W., Binder K., Heermann D. \and Kremer K.}
\REVIEW{J. Phys. II}{1}{1991}{37}.

\bibitem{com:entangled}
%In the following 
We describe as 'entangled-like' 
a system in which the diffusion constant follows a
$D\propto N^{-2}$ scaling, to emphasize that such 
a behaviour is not sufficient to assess the validity of 
the reptation picture.

\bibitem{Faller}
\Name{Faller R.}
preprint cond-mat/0005192.

\bibitem{Graessley}
\Name{Fetters L. J. \etal}
%, Lohse D. J., Milner S. T. \and Graessley W. W.}
\REVIEW{Macromolecules}{32}{1999}{6847}.

\bibitem{com:Fedderlength}
The so-called ``packing length" $p$ is sometimes prefered to the blob size
$\xi$ \cite{Graessley}.
Closer inspection shows that $p$ is essentially the same as $\xi$.
Following experiment\cite{Graessley} we have verified 
the scaling of fig.~\ref{figDscal} using $p$ instead of $\xi$.
This gives very similar results; 
the collapse with $\xi\sim\de$ being slightly better.

\bibitem{Lutz}
\Name{McKenna G. B. \etal}
% Hadziioannou G., Lutz P., Hild G., Strazielle C., 
%Straupe C., Rempp P. \and Kovacs A. J.}
\REVIEW{Macromolecules}{20}{1987}{498};
\Name{Ederl\'e Y. \etal}
%, Naraghi K. \and Lutz P.J.}
\Book{Materials Science and Technology, A Comprehensive Treatment}
\Editor{A. D. Schl\"uter}
\Vol{Synthesis of Polymers} 
\Publ{Wiley-VCH, Weinheim-New-York}\Year{1999}\Page{622}.

\bibitem{com:stiffness}
Obviously, this sets an upper bound $\bd(\sigma) \ll \xi(\sigma)$ 
on the stiffness variation.
For very large stiffness there are additional caveats: 
(i) The structure of the underlying monomer fluid changes with $\sigma$. 
This effect is partially captured by the stiffness dependence of the mobility $m$, 
but stiffness and density effects will not decouple for large $\sigma$. 
(ii) For even larger $\sigma$ the system must undergo a transition to a 
nematic structure. 
(iii) For large $\sigma$ and finite $N$ the large scale conformations differ from 
the properties of flexible molecules, i.e., chains become rods and rings beome 
circles.

\bibitem{Schaefer}
\Name{Ebert U. \etal}
%, Baumg\"artner A. \and Sch\"afer L.} 
\REVIEW{J.~Stat.~Phys.}{90}{1998}{1325};
%L.~Sch\"afer, A.~Baumg\"artner, U.Ebert, 
\REVIEW{European Phys. J. B}{10}{1999}{105}.

%\bibitem{inpreparation}
%\Name{M\"uller M., Wittmer J. P., Barrat J.-L. and Cates M.~E.}
%in preparation.

\bibitem{com:Rousemelt}
This clarifies various recent queries about the bad performance 
of the Rouse model in the melt limit at high frequencies. Strictly speaking, 
the Rouse description is only {\em marginally} valid in the melt 
limit $\xi \rightarrow 0$ and {\em no} proper Rouse plateau is expected there. 
This statement is in agreement with fig.~\ref{figDN}
and recent careful studies\cite{MWC96,paul,WPB92,papRousemelt} 
which clearly show that the forces on short chains in the melt are always correlated,
i.e. are {\em not} of Langevin type.
%
As soon as the chains overlap this must affect the dynamics because
$\de\sim\xi$; the effective forces on reference chains seen are, hence,
crossover effects. However, because $\de/\xi\gg 1$ it is allowed to view 
the Rouse assumption as a useful {\em approximation} for various, 
but not all observed quantities.
%
Note that the  experimental practice is to tune the dynamic behaviour of short
chains to the Rouse description by means of chain length dependent mobility
corrections \cite{lrp}. 
This may be misleading when assessing the validity of the Rouse model at small 
chain length in the melt limit.

\bibitem{papRousemelt}
\Name{Paul W., Smith G. D. \and Yoon D. Y.} 
\REVIEW{Macromolecules}{30}{1997}{7772}.
%\Name{Smith G. D., Paul W., Monkenbusch M., Willner L., 
%Richter D., Qiu X. H., Ediger M. D.}
%\REVIEW{Macromolecules}{32}{1999}{8857}.

\end{thebibliography}
\end{document}